\documentclass[letterpaper]{article} 
\usepackage{aaai2026}  
\usepackage{times}  
\usepackage{helvet}  
\usepackage{courier}  
\usepackage[hyphens]{url}  
\usepackage{graphicx} 
\usepackage{natbib}  
\usepackage{caption} 
\usepackage{algorithm}
\usepackage{algorithmic}
\usepackage{booktabs}
\usepackage{diagbox}
\usepackage{multirow}
\usepackage{diagbox}
\usepackage{makecell} 
\usepackage{amsmath} 
\usepackage{amssymb}
\usepackage{subcaption} 
\usepackage{appendix}
\usepackage[font=small,labelfont=bf]{caption}  
\usepackage{newfloat}
\usepackage{listings}
\usepackage{bm}
\DeclareCaptionStyle{ruled}{labelfont=normalfont,labelsep=colon,strut=off} 
\floatstyle{ruled}
\newfloat{listing}{tb}{lst}{}
\floatname{listing}{Listing}
\title{UMRE: A Unified Monotonic Transformation for Ranking Ensemble in Recommender Systems}
\author{
\begin{tabular}[t]{ccc}
\parbox{0.3\linewidth}{\centering
\textbf{\large Zhengrui Xu\footnotemark[2]}\\
\textmd{\normalsize Beijing Jiaotong University}\\
\textmd{\normalsize Beijing, China}\\
\textmd{\normalsize zrxu23@bjtu.edu.cn}
}
&
\parbox{0.3\linewidth}{\centering
\textbf{\large Zhe Yang\footnotemark[2]}\\
\textmd{\normalsize Kuaishou Technology}\\
\textmd{\normalsize Beijing, China}\\
\textmd{\normalsize yangzhe03@kuaishou.com}
}
&
\parbox{0.3\linewidth}{\centering
\textbf{\large Zhengxiao Guo}\\
\textmd{\normalsize Kuaishou Technology}\\
\textmd{\normalsize Beijing, China}\\
\textmd{\normalsize guozhengxiao@kuaishou.com}
}
\\[2em]
\parbox{0.3\linewidth}{\centering
\textbf{\large Shukai Liu}\\
\textmd{\normalsize Kuaishou Technology}\\
\textmd{\normalsize Beijing, China}\\
\textmd{\normalsize shukailiu89@gmail.com}
}
&
\parbox{0.3\linewidth}{\centering
\textbf{\large Luocheng Lin}\\
\textmd{\normalsize Kuaishou Technology}\\
\textmd{\normalsize Beijing, China}\\
\textmd{\normalsize 21210180058@m.fudan.edu.cn}
}
&
\parbox{0.3\linewidth}{\centering
\textbf{\large Xiaoyan Liu}\\
\textmd{\normalsize Kuaishou Technology}\\
\textmd{\normalsize Beijing, China}\\
\textmd{\normalsize liuxiaoyan18@mails.ucas.ac.cn}
}
\\[2em]
\multicolumn{3}{c}{
\begin{tabular}{cc}
\parbox{0.3\linewidth}{\centering
\textbf{\large Yongqi Liu\footnotemark[1]}\\
\textmd{\normalsize Kuaishou Technology}\\
\textmd{\normalsize Beijing, China}\\
\textmd{\normalsize liuyongqi@kuaishou.com}
}
&
\parbox{0.3\linewidth}{\centering
\textbf{\large Han Li}\\
\textmd{\normalsize Kuaishou Technology}\\
\textmd{\normalsize Beijing, China}\\
\textmd{\normalsize lihan08@kuaishou.com}
}
\end{tabular}
}
\end{tabular}
}

\usepackage{bibentry}

\begin{document}

\maketitle
\footnotetext[2]{Equal Contribution.} 
\footnotetext[1]{Corresponding Author.} 

\begin{abstract}
Industrial recommender systems commonly rely on ensemble sorting (ES) to combine predictions from multiple behavioral objectives. Traditionally, this process depends on manually designed nonlinear transformations (e.g., polynomial or exponential functions) and hand-tuned fusion weights to balance competing goals—an approach that is labor-intensive and frequently suboptimal in achieving Pareto efficiency. In this paper, we propose a novel \textbf{U}nified \textbf{M}onotonic \textbf{R}anking \textbf{E}nsemble (\textbf{UMRE}) framework to address the limitations of traditional methods in ensemble sorting. UMRE replaces handcrafted transformations with Unconstrained Monotonic Neural Networks (UMNN), which learn expressive strictly monotonic functions through the integration of positive neural integrals. Subsequently, a lightweight ranking model is employed to fuse the prediction scores, assigning personalized weights to each prediction objective. To balance competing goals, we further introduce a Pareto optimality strategy that adaptively coordinates task weights during training. UMRE eliminates manual tuning, maintains ranking consistency, and achieves fine-grained personalization. Experimental results on two public recommendation datasets (Kuairand, Tenrec) and online A/B tests demonstrate impressive performance and generalization capabilities.

\end{abstract}


\section{Introduction}

Recommender systems play a crucial role across a wide range of platforms, including e-commerce~\cite{gu2020hierarchical,linden2003amazon,zhou2018deep}, videos~\cite{tang2017popularity,wu2018beyond}, and news~\cite{liu2010personalized, zheng2018drn}. In many real-world scenarios, users generate multiple types of behavioral feedback within a single session. For example, on video platforms, users may click, like, share, or follow content. Modeling such diverse user behaviors is essential for improving recommendation quality and user satisfaction.

\begin{figure}[htbp]
  \centering
  \includegraphics[width=0.9\linewidth]{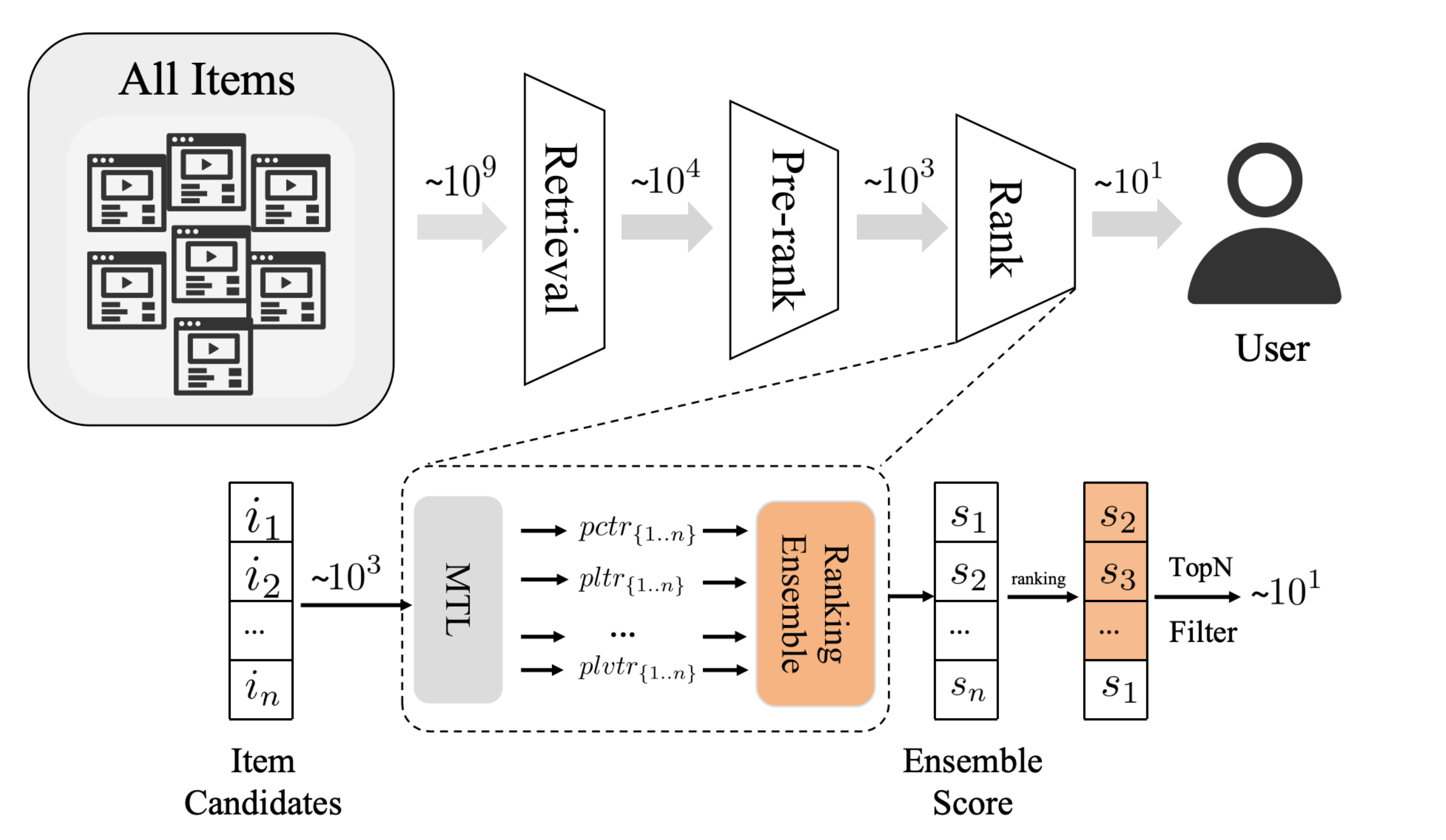}
  \caption{Funnel-shaped architecture for recommendation with a two-stage ranking framework. This paper focuses on the research of the ranking ensemble stage}
  \label{fig:recommendation_funnel}
\end{figure}

Most industrial recommender systems adopt a multi-stage pipeline based on the retrieval–ranking paradigm, where ranking is typically divided into pre-rank and rank. At each stage, two components are involved: (1) a multi-task learning (MTL) model that predicts multiple user behavior probabilities (e.g., click-through rate, like rate), and (2) ensemble sorting that combines these predictions into a ensemble score. The top-ranked items based on this score are passed to the next stage. Figure~\ref{fig:recommendation_funnel} illustrates the overall architecture.

We denote the predicted probability of behavior “x” as pxtr (e.g., pctr for clicks, pltr for likes). MTL methods such as MMoE~\cite{ma2018modeling} and PLE~\cite{tang2020progressive} have shown strong performance in jointly modeling multiple tasks using shared experts and gating mechanisms over user, item, and interaction features. However, a user’s overall interest in an item is rarely captured by a single pxtr. To address this, a ranking ensemble framework is used to combine multiple pxtrs into a single ensemble score that reflects overall user preference. This ranking ensemble process typically includes two steps: pxtr transformation, which reshapes the distribution of pxtrs (e.g., using scaling factors to enhance score separability), and pxtr fusion, which aggregates them into a final ranking score.

In practice, most systems rely on predefined fusion rules, such as weighted sums or products. These methods are efficient and easy to deploy, allowing for quick online adjustments without retraining. However, they require extensive manual tuning, especially when many behavioral signals are involved. Moreover, static weights fail to capture individual user preferences, thereby limiting personalization.

Recent studies~\cite{zhang2025embed,meng2025generative,li2023intent} have investigated model-based fusion, yet most focus solely on the fusion step while overlooking the transformation stage, thereby limiting the optimization space. A further challenge is the lack of explicit supervision for the ensemble task. Unlike multi-task learning (MTL), which benefits from well-defined binary labels (e.g., "like" vs. "dislike"), ranking ensemble models often lack direct ground-truth signals to guide the fusion process. (1) A common practice is to treat a primary objective—such as watch time or long view rate—as the sole supervision signal. However, this reduces the learning signal to a single behavioral metric, potentially neglecting other informative pxtrs and valuable user feedback. (2) Another line of work adopts reinforcement learning~\cite{cai2023reinforcing, chen2024cache, liu2024off, zhang2024unex, zhang2022multi}, using real-time engagement metrics as rewards. Yet in mature platforms, multiple conflicting metrics coexist, and no single one fully captures user satisfaction. (3) A more principled solution is to construct a composite reward by weighting and aggregating multiple behavior labels, but determining optimal weights remains challenging. Manual tuning is labor-intensive and often suboptimal—particularly in dynamic environments where user preferences and platform goals continually shift.

\begin{table}[t]
\centering
\caption{Typical fusion formulas in existing research.}
\renewcommand{\arraystretch}{1.8} 
\label{table: Typical fusion formulas}
\begin{tabular}{c|c|c}
\hline
\hline
\textbf{Index} & \textbf{Type} & \textbf{Formula} \\
\hline
1 & Additive &  $s_i =  {\textstyle \sum_{k=1}^{K}}w_k (\alpha _k p_{ki} + b_k)^{\beta_k }$\\
\hline
2 & Multiplicative& $s_i =   {\textstyle \prod_{k=1}^{K}}((\alpha _k p_{ki} + b_k)^{\beta_k })^{w_k}$\\
\hline
\hline
\end{tabular}
\end{table}

To address these challenges, we propose UMRE (Unified Monotonic Ranking Ensemble), an end-to-end model that performs personalized pxtr transformation and fusion. First, we obtain task-specific pxtrs from a pretrained MTL model. These are passed through a UMNN (Unconstrained Monotonic Neural Network), which guarantees monotonic transformation while learning user-specific scaling functions. The transformed pxtrs, combined with user history and item features, are then input into a fusion model. We further introduce a Pareto-optimal optimization strategy that adjusts task weights during training based on changes in evaluation metrics, achieving better multi-objective trade-offs. All modules are trained jointly to maximize overall performance and personalization. Our contributions are summarized as follows:
\begin{enumerate}
\item We propose the UMRE model, a novel end-to-end ranked ensemble model that personalises the fusion of multiple task predictions.

\item We have introduced the pxtrs monotonic transformation based on UMNN, which preserves its relative order while reshaping the pxtr distribution and enabling user-specific scaling.

\item We design a Pareto-optimal optimization strategy, which dynamically adjusts multi-task weights during training, eliminating manual tuning and facilitating balanced optimization.

\item Our method was validated in two public recommendation datasets and online A/B testing. UMRE achieves optimal performance on all tasks compared to other baselines and has achieved significant benefits on online platforms.
\end{enumerate}

\section{Related Work}

\subsection{Multi-Task Learning}

In industrial recommender systems, it is common to predict multiple types of user feedback for each candidate item. For example, in e-commerce platforms, the system must not only recommend items of interest but also encourage downstream actions such as conversions or purchases. This requires the joint modeling of multiple objectives, such as click-through rate (CTR) and conversion rate (CVR).

Neural network-based multi-task learning (MTL)~\cite{chen2018gradnorm,ma2018entire,ma2018modeling,tang2020progressive,su2024stem,yang2023adatask,yu2020gradient} has become a standard approach for addressing this challenge. A notable example is ESMM from Alibaba, which adopts a shared-bottom architecture to jointly model CTR and CVR. This design effectively mitigates issues like data sparsity and sample selection bias in CVR prediction. Later, MMoE introduced by Google extends this idea with a mixture-of-experts framework and task-specific gating, enabling better balance between shared knowledge and task specialization.

Building on this, PLE~\cite{tang2020progressive} proposed a layered expert structure that explicitly separates shared and task-specific representations, significantly alleviating negative transfer and improving performance in large-scale systems. More recent MTL approaches incorporate techniques such as user intent modeling, adversarial disentanglement, and contrastive learning to further enhance representation learning and task-level generalization.

In summary, MTL has proven to be a robust and essential paradigm for simultaneously optimizing multiple objectives in industrial recommendation scenarios.

\subsection{Ranking Ensemble}
Following the MTL stage, where each task outputs a distinct prediction score (pxtr) for candidate items, a ranking ensemble mechanism integrates these signals to produce the final ranked list for item exposure. Existing ensemble methods can be broadly categorized into two paradigms: predefined formula-based and learning-based ensemble models.

In the formula-based paradigm, task scores are combined using fixed mathematical functions—typically additive or multiplicative forms—as illustrated in Table~\ref{table: Typical fusion formulas}(1)(2). Task importance is reflected by assigning weights to each score. However, these weights are often manually tuned, and the hyperparameter space grows combinatorially with the number of tasks, making optimization difficult. To mitigate this, reinforcement learning (RL)~\cite{sutton1998reinforcement}-based methods~\cite{rubinstein2004cross} treat score weights as actions and optimize them through interaction with user environments. Despite this, most RL-based approaches lack user-level personalization, limiting their flexibility in diverse scenarios.

Learning-based ensemble methods, in contrast, model the fusion process via supervised learning. For instance, Oliveira et al.~\cite{oliveira2016evolutionary} proposed Evolutionary Rank Aggregation (ERA) using genetic programming, while Bałchanowski and Boryczka~\cite{balchanowski2022aggregation,balchanowski2022collaborative} applied differential evolution. Zhang et al.~\cite{zhang2022multi} employed RL for rank fusion, and other works~\cite{li2023intent} leverage user behavior and contextual signals to enable intent-aware personalized fusion. He et al.~\cite{he2025end} further introduced a Pareto-based self-evolutionary framework to achieve personalized fusion, which adaptively balances multiple optimization objectives during the aggregation process.

Overall, this evolution reflects a shift from rule-based heuristics to data-driven, learnable fusion strategies, emphasizing adaptability and personalization in ranking systems.

\subsection{Pxtr Transformation}
Before ranking ensemble, task-specific predictions (pxtr) are typically transformed to enhance discriminability and align score distributions across tasks. This step ensures that no single task dominates due to inherently larger score magnitudes, while improving intra-task ranking quality.

Traditional methods apply polynomial transformations, adjusting parameters such as scaling factors $\alpha$, $b$, and $\beta$, as shown in Table~\ref{table: Typical fusion formulas}. However, as the number of tasks increases, the hyperparameter space grows rapidly, making tuning both computationally expensive and inefficient.

To address this, recent work explores neural network-based transformations using MLPs~\cite{cao2025xmtf}, which learn flexible nonlinear mappings from pxtr scores. While expressive, these methods lack monotonicity guarantees, potentially distorting score order and undermining intra-task ranking fidelity.

To overcome this limitation, we adopt Unconstrained Monotonic Neural Networks (UMNNs)~\cite{wehenkel2019unconstrained} to model score transformations. By formulating the transformation as an integral over a learned non-negative function, UMNNs ensure strict monotonicity. Moreover, by conditioning the integrand on user and item features, the transformation becomes adaptive and personalized. This approach preserves relative ranking within tasks while harmonizing score scales across tasks, enhancing the effectiveness of subsequent ranking ensemble methods.

\begin{figure*}[htbp]
\vspace{-2em}
    \centering
    \includegraphics[width=0.95\textwidth]{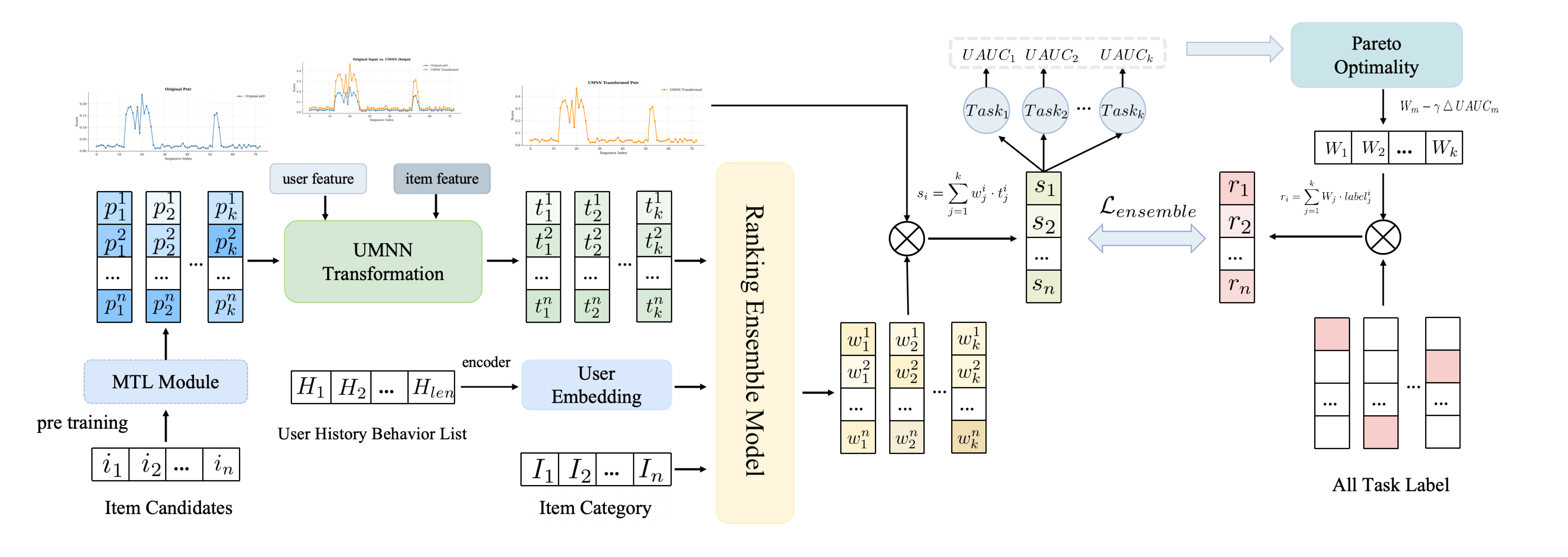}
    \caption{An overview of the proposed UMRE pipeline. We adopt Progressive Layered Extraction (PLE) as the pre-trained multi-task learning (MTL) model, employ GRU4Rec as the sequence encoder, and define the ensemble loss $\mathcal{L}_{\text{ensemble}}$ as shown in Equation~\ref{equation:loss}.}
    \label{fig:pipeline}
\end{figure*}

\section{METHODOLOGY}

\subsection{Problem Statement}
In recommendation systems, as summarized in Table~\ref{table:Key Notations}, let \(\mathcal{U}\) denote the user set and \(\mathcal{I}\) the item set. For each user-item pair \((u,i) \in \mathcal{U} \times \mathcal{I}\), the ranking stage produces \(K\) distinct prediction targets (e.g., like rate, follow rate) denoted as the prediction vector:
\[
\mathbf{p} = (p_1, p_2, \ldots, p_K) \in \mathbb{R}^K
\]
where each \(p_k\) represents the prediction for objective \(k\) (abbreviated as \textit{pxtr}).

The final ranking requires a \textbf{ensemble score} \(s(u,i)\) derived through \textit{Ensemble Sorting (ES)}. This process involves:
    

\noindent\textbf{Pxtr Transformation}:
Apply transformation functions \(g_k: \mathbb{R} \to \mathbb{R}\) to each pxtr:
\[
t_k = g_k(p_k), \quad \forall k \in \{1,\ldots,K\}
\]
To preserve the ranking order implied by the original pxtr, the function \(g_k\) must be \textit{monotonic}:
\[
p_k^{(a)} > p_k^{(b)} \implies g_k(p_k^{(a)}) > g_k(p_k^{(b)})
\]


\noindent\textbf{Pxtr Fusion}:
Combine transformed outputs using a function \(F\) with weight parameters \(\mathbf{w}\):
\[
s = F(\mathbf{t}, \mathbf{w}), \quad \mathbf{t} = (t_1, \ldots, t_K)
\]
The objective of Ensemble Sorting is to jointly learn transformation functions \(g_k\) and fusion function \(F\) such that the resulting ensemble score yields a desirable ranking—achieving \textit{Pareto optimality} across multiple objectives.
Existing formula-based fusion methods suffer from fixed weight assignments, limited personalization, and poor scalability to many tasks. In addition, the common two-stage approach—separately optimizing pxtr transformation and fusion—fails to reach global optima.


\begin{table}[h]
\centering
\caption{Notations used in this paper}
\label{table:Key Notations}
\begin{tabular}{cl}
\hline
\multicolumn{1}{c}{\textbf{Symbol}} & \multicolumn{1}{c}{\textbf{Definition}} \\
\hline
\(u \in \mathcal{U}\) & User in user set \\
\(i \in \mathcal{I}\) & Item in item set \\
\(p_k(u,i)\) & Raw prediction score for objective \(k\) \\
\(g_k(\cdot)\) & Monotonic transformation function \\
\(t_k(u,i)\) & Transformed prediction score \\
\(F(\cdot)\) & Fusion function (e.g., linear weighting) \\
\(s(u,i)\) & Final ensemble score \\
\(\theta_k\) & Parameters of handcrafted \(g_k\) \\
\(\mathbf{w}\) & Fusion weights \\
\hline
\end{tabular}
\end{table}

\subsection{Overall Framework}

To overcome the limitations of manual tuning and poor generalization in traditional Ensemble Sorting, we propose a fully learnable end-to-end framework that replaces both the transformation functions \(g_k\) and the fusion function \(F\) with trainable neural components.



As illustrated in Figure~\ref{fig:pipeline}, we first apply \textit{Unconstrained Monotonic Neural Networks (UMNNs)} to model each transformation function \(g_k\), which guarantees strict monotonicity while enabling personalized and non-linear scaling of each \textit{pxtr} score \(p_k\). The transformed score is defined as:
\begin{equation}
t_k = g_k(p_k) = \text{UMNN}(\text{pxtr}_k)
\end{equation}

Subsequently, user and item representations are derived from historical behavior sequences and item metadata. A \textit{cross-attention mechanism} is then employed to compute dynamic fusion weights:
\[
\mathbf{w} = (w_1, w_2, \ldots, w_K),
\]

These weights are used to compute the final ensemble score via weighted summation:
\begin{equation}
s = \sum_{k=1}^K w_k \cdot t_k
\end{equation}

The model is trained end-to-end using \textit{mean squared error (MSE)} against a global reward signal \(r\), which integrates multiple user feedback types:
\begin{equation}
r = \sum_{k \in K} W_k \cdot g_k
\end{equation}
where  \(W_k\) is the importance weight for behavior \(k\), and \(g_k\) is its binary label.

Importantly, the behavior weights \(W_k\) are not static. They are dynamically updated during training through \textit{Pareto optimization}, which jointly considers:
\begin{itemize}
    \item Balancing optimization across multiple objectives
    \item Enabling autonomous model improvement
\end{itemize}

This framework eliminates manual score engineering by integrating pxtr transformation and pxtr fusion into a unified model, enabling end-to-end training for personalized, \textit{Pareto-optimal} ensemble ranking across multiple recommendation objectives.

\subsection{UMNN Transformation Module}

To model the monotonic transformation functions \(g_k\) for each prediction target \(p_k\), we adopt \textit{Unconstrained Monotonic Neural Networks (UMNN)}~\cite{wehenkel2019unconstrained}. Compared with handcrafted transformations, UMNN offer three key advantages:  
\begin{itemize}
    \item High function expressiveness without structural constraints.
    \item Theoretical guarantees of strict monotonicity.
    \item Supports personalised non-linear scaling by user.
\end{itemize}

\begin{figure}[htbp]
  \centering
  \includegraphics[width=0.9\linewidth]{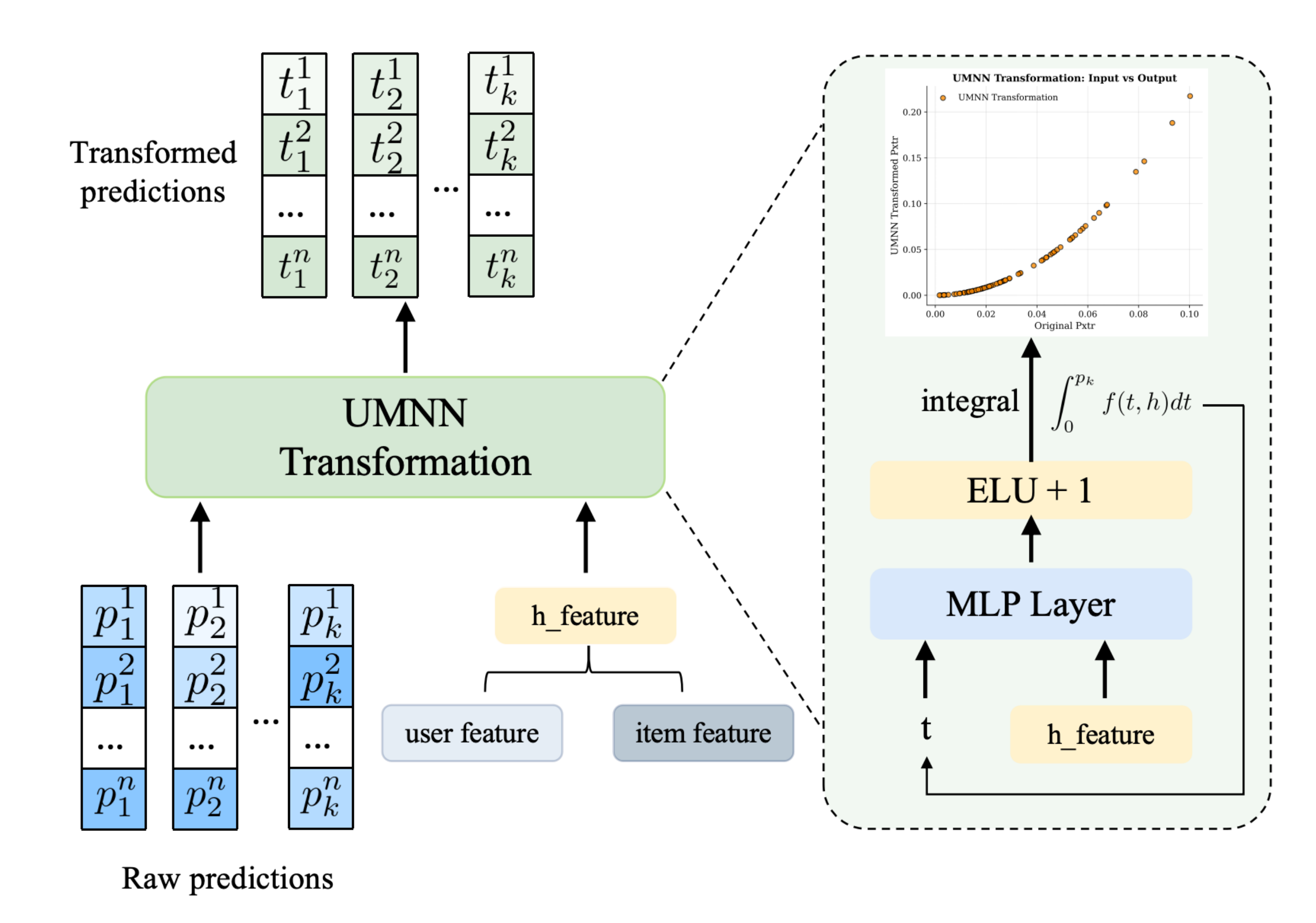}

  \caption{Structure of the UMNN module.}
  \label{fig:UMNN Module}

\end{figure}

\noindent \textbf{Integral Function}: As shown in Figure~\ref{fig:UMNN Module}, we define the integral function \(f_k(t, h; \theta_k)\), which acts as the integrand in the transformation. This function can adopt any deep neural network architecture, in this work, we employ a multi-layer perceptron (MLP~\cite{rosenblatt1958perceptron}). The inputs to \(f_k\) include the integral variable \(t\) and a personalized feature vector \(h\) (e.g., user and item embeddings), enabling the function to adapt to different user profiles. To ensure the strict monotonicity of the resulting transformation \(g_k\), we constrain the output of \(f_k\) to be strictly positive by applying an ELU+1 activation function:
\begin{equation}
f_k(t, h; \theta_k)) = \text{ELU}(\text{MLP}(t,h)) + 1
\end{equation}
\noindent \textbf{Monotonic Integral Transformation}:  
Each transformation \(g_k(\cdot)\) is defined as:
\begin{equation}
t_k = g_k(p_k, h; \theta_k) = \int_0^{p_k} f_k(t, h; \theta_k) \, dt + \beta_k
\end{equation}
where:
\begin{itemize}
    \item \(f_k(t, h; \theta_k)\) is an unconstrained neural network with strictly positive outputs, enforced via an ELU+1 activation function;
    \item \(h\) is the feature vector containing the user embedding, item embedding;
    \item \(\theta_k\) denotes the parameters specific to the \(k\)-th transformation network;
    \item \(\beta_k\) is a learnable bias term.
\end{itemize}
The strict positivity constraint on \(f_k\) guarantees the monotonicity of \(g_k\), ensuring:
\[
p_k^{(a)} > p_k^{(b)} \implies g_k(p_k^{(a)}, h) > g_k(p_k^{(b)}, h)
\]
which is critical for preserving the relative order of prediction targets in ranking applications.



\noindent More detailed description can be found in Appendix A.

\noindent \textbf{Output:} Transformed pxtrs \(\mathbf{t} = (t_1, \ldots, t_K)\) serve as input to the ranking ensemble module.

\begin{figure}[htbp]
  \centering
  \includegraphics[width=0.9\linewidth]{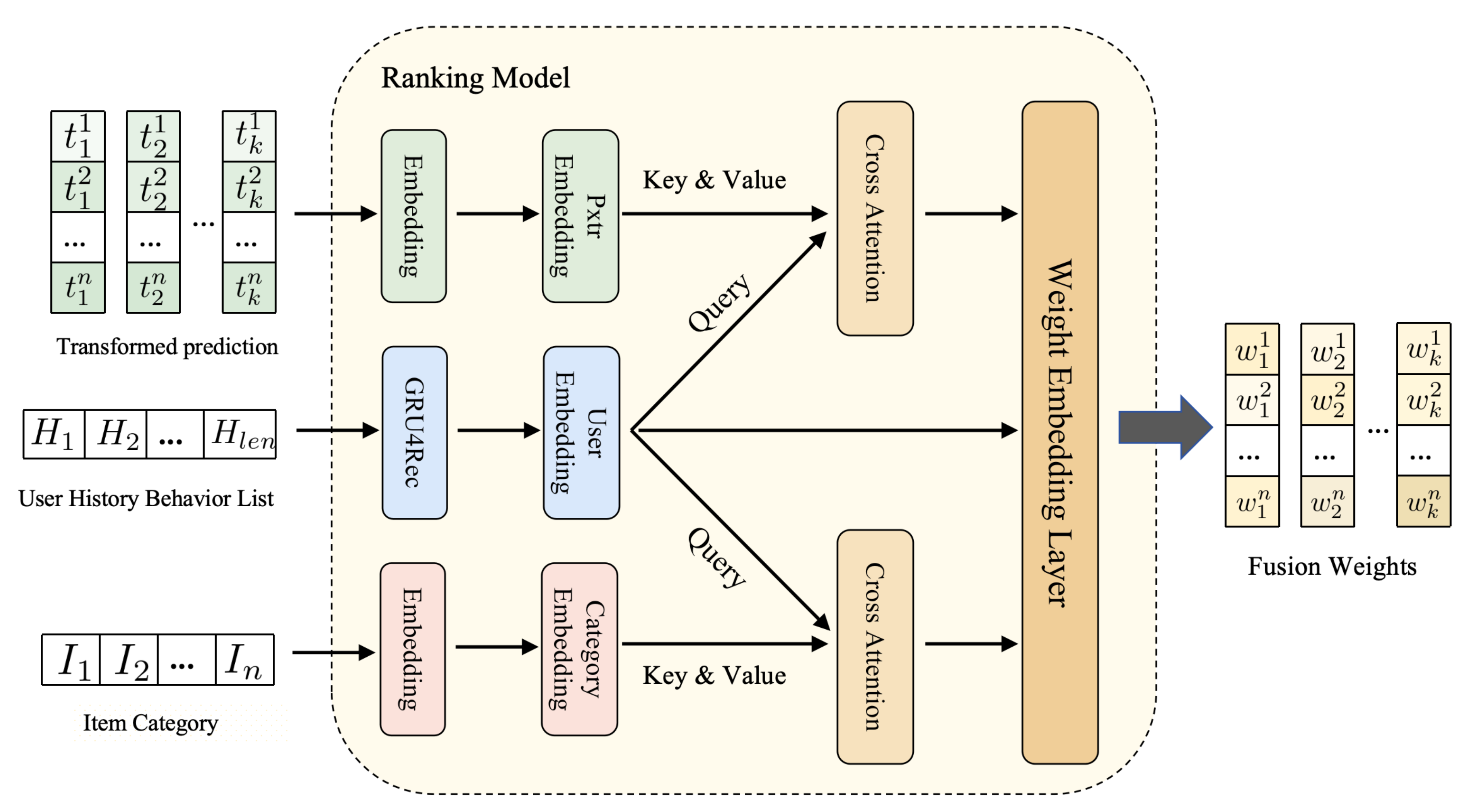}

  \caption{Structure of the Ranking Ensemble Model.}
  \label{fig:Ranking_Model}

\end{figure}

\subsection{Ranking Ensemble Module}

The Ranking Ensemble Module adaptively integrates transformed predictions \(\mathbf{t} = (t_1, \dots, t_K)\) through a context-aware attention mechanism, learning optimal fusion weights \(\mathbf{w} = (w_1, \dots, w_K)\). This approach replaces manual hyperparameter tuning with a learnable, personalized weighting strategy that dynamically reflects user intent. As shown in Figure~\ref{fig:Ranking_Model}, the module comprises three components:

\noindent \textbf{Encoding of User Behavior Sequences.}   
Each user's interaction history is first encoded as:
\[
\mathbf{H} = \left[ \mathbf{e}_1 \oplus \mathbf{e}_c^1 \oplus \mathbf{e}_a^1 ; \cdots ; \mathbf{e}_T \oplus \mathbf{e}_c^T \oplus \mathbf{e}_a^T \right]
\]
where \(\mathbf{e}_t\), \(\mathbf{e}_c^t\), and \(\mathbf{e}_a^t\) denote the item, category, and action-type embeddings at timestamp \(t\), respectively, and \(\oplus\) denotes concatenation. The concatenated representation captures the temporal evolution and semantic diversity of user intent across \(T\) historical interactions. 

The sequence \(\mathbf{H}\) is input into a GRU4Rec~\cite{hidasi2015session} encoder to produce a user embedding (\( \mathbf{U_{\text{emb}}} \)):
\[
\mathbf{U_{\text{emb}}} = \text{GRU4Rec}(\mathbf{H}).
\]

\noindent \textbf{Cross-Attention-Based Weight Learning.}  
The learned user embedding \(\mathbf{u}\) is then used as the query to perform cross-attention with both the transformed prediction embeddings \(\mathbf{T}_{\text{emb}}\) and the category embeddings \(\mathbf{C}_{\text{emb}}\):
\begin{align*}
\mathbf{Q}_1 &= \mathbf{U_{\text{emb}}}\mathbf{W}_q^1, \quad \mathbf{K}_1 = \mathbf{T}_{\text{emb}}\mathbf{W}_k^1, \quad \mathbf{V}_1 = \mathbf{T}_{\text{emb}}\mathbf{W}_v^1, \\
\mathbf{Q}_2 &= \mathbf{U_{\text{emb}}}\mathbf{W}_q^2, \quad \mathbf{K}_2 = \mathbf{C}_{\text{emb}}\mathbf{W}_k^2, \quad \mathbf{V}_2 = \mathbf{C}_{\text{emb}}\mathbf{W}_v^2.
\end{align*}
The two attention outputs are computed as:
\begin{align*}
\mathbf{A}_1 &= \text{softmax}\left( \frac{\mathbf{Q}_1 \mathbf{K}_1^\top}{\sqrt{d_k}} \right) \mathbf{V}_1, \\
\mathbf{A}_2 &= \text{softmax}\left( \frac{\mathbf{Q}_2 \mathbf{K}_2^\top}{\sqrt{d_k}} \right) \mathbf{V}_2.
\end{align*}
The outputs \(\mathbf{A}_1\) and \(\mathbf{A}_2\) are concatenated together with the user embedding \(\mathbf{U}_{\text{emb}}\), and then passed through weight embedding layer to produce the fusion weights:
\[
\mathbf{w} = \left( \text{Linear}\left( \mathbf{U_{\text{emb}}} \oplus \mathbf{A}_1 \oplus \mathbf{A}_2 \right) \right),
\]

\noindent \textbf{Adaptive Fusion of Predictions.}  
The final ensemble score is computed as a weighted sum of the transformed predictions:
\begin{equation}
s = \sum_{k=1}^K w_k \cdot t_k.
\end{equation}
This ensemble sorting inherits the monotonicity guarantees of UMNN while enabling context-aware, objective-specific reweighting. The model is trained end-to-end by minimizing the mean squared error (MSE) between the fused score \(s\) and the reward signal \(r\):
\begin{equation}
\label{equation:loss}
\mathcal{L}_{\text{ensemble}} = \frac{1}{N} \sum_{i=1}^N \left( s^{(i)} - r^{(i)} \right)^2.
\end{equation}

\subsection{Pareto-Optimal Reward Design}

In order to balance competing objectives, we propose a dynamic reward mechanism that can be adjusted throughout the training process to achieve a Pareto-optimal trade-off between multiple user behaviours.

\noindent \textbf{Reward Signal Construction.}
We define a weighted composite reward over a set of engagement indicators \(\mathcal{M}\):
\begin{equation}
r_{init} = \sum_{m \in \mathcal{M}} \omega_m \cdot y_m, \quad \text{s.t.} \quad \sum_{m} \omega_m = 1
\end{equation}
where \(y_m \in \{0,1\}\) denotes the binary feedback for metric \(m\), and \(\omega_m\) represents its relative importance. The weights can be initialised before training.

\noindent \textbf{Adaptive Weight Optimization.}
To approach Pareto-optimality, we employ an epoch-wise adjustment strategy based on changes in UAUC (User-level AUC). The procedure is summarized below:

\begin{algorithm}[H]
\caption{Pareto Reward Optimization}
\begin{algorithmic}[1]
\STATE Initialize weights \(\bm{\omega}^{(0)}\), epoch \(e \gets 0\)
\REPEAT
    \STATE Train model with reward \(r^{(e)} = \mathbf{y} \cdot \bm{\omega}^{(e)}\)
    \STATE Evaluate UAUC: \(\mathbf{u}^{(e)} = [\text{UAUC}_1^{(e)}, \dots, \text{UAUC}_5^{(e)}]\)
    \IF{$e \geq E_s$}
        \STATE \(\delta_m \gets \gamma \cdot (\text{UAUC}_m^{(e-1)} - \text{UAUC}_m^{(e)})\)
        \STATE \(\omega_m^{(e+1)} \gets \frac{\omega_m^{(e)} - \delta_m}{\sum_k (\omega_k^{(e)} - \delta_k)}\)
        \STATE Clip: \(\omega_m^{(e+1)} \leftarrow \text{clip}_{[\omega_{\min}, \omega_{\max}]}(\omega_m^{(e+1)})\)
    \ENDIF
    \STATE \(e \gets e + 1\)
\UNTIL{\(\|\bm{\omega}^{(e)} - \bm{\omega}^{(e-1)}\|_1 < eps\) or \(e = E_{\max}\)}
\end{algorithmic}
\end{algorithm}

Let \(\bm{\omega}^{(e)}\) denote the reward weights at epoch \(e\), with training reward \(r^{(e)} = \mathbf{y} \cdot \bm{\omega}^{(e)}\), and evaluation metrics \(\mathbf{u}^{(e)} = [\text{UAUC}_1^{(e)}, \dots, \text{UAUC}_M^{(e)}]\). After a warm-up of \(E_s\) epochs, weights are updated using degradation \(\delta_m\) and step size \(\gamma\), then clipped to \([\omega_{\min}, \omega_{\max}]\) and normalized.

\noindent \textbf{Optimization Dynamics.}
The weight adjustment induces a negative feedback mechanism:
\begin{equation}
\omega_m^{(e+1)} \propto \omega_m^{(e)} - \gamma \cdot \Delta\text{UAUC}_m
\end{equation}
with the following interpretation:
\begin{itemize}
    \item \(\text{UAUC}_m \downarrow \Rightarrow \delta_m > 0 \Rightarrow\) \textbf{Increase} \(\omega_m\): prioritize underperforming metrics.
    \item \(\text{UAUC}_m \uparrow \Rightarrow \delta_m < 0 \Rightarrow\) \textbf{Decrease} \(\omega_m\): reallocate focus elsewhere.
\end{itemize}

\noindent This adaptive reward shaping drives the system towards a Pareto-efficient mechanism. Unlike static heuristics, our approach enables dynamic adjustment of weights throughout training, allowing real-time responsiveness to the optimization status of each objective. This facilitates effective coordination and balance among multiple goals, thereby enhancing overall recommendation performance.

\section{Experiments}
\subsection{Experimental Setup}
\subsubsection{Datasets and Evaluation metrics.}The experiments are conducted on two publicly available datasets: the KuaiRand~\cite{gao2022kuairand} dataset for online short video recommendation, and the Tenrec~\cite{yuan2022tenrec} dataset for online shopping recommendation. Detailed statistics of the dataset can be found in Table~\ref{tab:dataset-stats} and Appendix B.

For evaluation, we compute HR(Hit Rate) and NDCG for each task label based on the ensemble score.

\begin{table}[ht]
\centering
\caption{Statistics of datasets}
\label{tab:dataset-stats}
\begin{tabular}{lrr}
\toprule
\textbf{Datasets} & \textbf{KuaiRand} & \textbf{Tenrec} \\
\midrule
\#user       & 1,000       & 50,000     \\
\#item       & 4,369,953   & 1,559,905   \\
\#exposure   & 11,713,045  & 31,701,064 \\
\#click      & 4,429,840   & 6,024,518  \\
\#long view  & 3,069,461   & -           \\
\#like       & 182,842     & 337,201   \\
\#follow     & 11,398      & 23,115     \\
\#comment    & 31,126      & -           \\
\#forward    & 9,191       & 40,498     \\
\bottomrule
\end{tabular}
\end{table}

\begin{table*}[htbp]
\centering
\renewcommand{\arraystretch}{1.2} 
\caption{Comparison of baseline models on the KuaiRand and Tenrec dataset across multiple tasks using HR@3 and NDCG@3. }
\resizebox{\textwidth}{!}{
\begin{tabular}{c|c|cccccc|cccccc}
\hline
\makecell{Data-} & \multirow{2}{*}{\diagbox{Model}{Metric}} & \multicolumn{6}{c|}{HR@3} & \multicolumn{6}{c}{NDCG@3} \\
set &  & Click & Like & Follow & Comment & Forward & Long view & Click & Like & Follow & Comment & Forward & Long view \\
\hline

\multirow{6}{*}{\rotatebox{90}{KuaiRand}} 
& SingleSort & 0.8418\textsuperscript{*} & 0.3296\textsuperscript{*} & 0.1022\textsuperscript{*} & 0.1198\textsuperscript{*} & 0.1046\textsuperscript{*} & 0.7482\textsuperscript{*} & 0.5529\textsuperscript{*} & 0.1889\textsuperscript{*} & 0.0534\textsuperscript{*} & 0.0627\textsuperscript{*} & 0.0580\textsuperscript{*} & 0.4461\textsuperscript{*}\\
& LR & 0.8485 & \underline{0.2818} & 0.0784 & 0.0868 & 0.0663 & 0.7615 & 0.5841 & \underline{0.1450} & 0.0437 & 0.0476 & 0.0363 & 0.5034\\
& MLP & 0.8477 & 0.2664 & 0.0902 & 0.0932 & \underline{0.0673} & 0.7772 & 0.5731 & 0.1289 & \underline{0.0612} & 0.0474 & 0.0384 & 0.4902\\
& aWELv & 0.8634 & 0.2041 & 0.0640 & \underline{0.1289} & 0.0635 & 0.7710 & 0.5654 & 0.1063 & 0.0356 & \underline{0.0652} & \underline{0.0565} & 0.4622\\
& IntEL& \underline{0.8821} & 0.2490 & \underline{0.1163} & 0.0792 & 0.0335 & \underline{0.8118} & \underline{0.6064} & 0.1255 & 0.0546 & 0.0422 & 0.0135 & \underline{0.5197}\\
& \textbf{UMRE} & \textbf{0.9523} & \textbf{0.4629}& \textbf{0.3502} & \textbf{0.2653} & \textbf{0.2113} & \textbf{0.9192} & \textbf{0.7712} & \textbf{0.3145} & \textbf{0.2693} & \textbf{0.1648} & \textbf{0.1393} & \textbf{0.7074}\\
\hline

\multirow{6}{*}{\rotatebox{90}{Tenrec}} 
& SingleSort & 0.8512\textsuperscript{*} & 0.4852\textsuperscript{*} & 0.4191\textsuperscript{*} & -- & 0.6549\textsuperscript{*} & -- & 0.5855\textsuperscript{*} & 0.3133\textsuperscript{*} & 0.3256\textsuperscript{*} & -- & 0.5399\textsuperscript{*} & --\\
& LR & 0.8404 & 0.4754 & 0.3844 & -- & 0.6418 & -- & 0.5750 & 0.3077 & 0.3049 & -- & 0.5246 & --\\
& MLP & 0.8455 & 0.4704 & 0.3932 & -- & 0.6400 & -- & \underline{0.5796} & 0.3051 & 0.3135 & -- & 0.5275 & --\\
& aWELv & 0.8274 & \underline{0.4772 }& \underline{0.4081} & -- & \underline{0.6514} & -- & 0.5595 & \underline{0.3101} & \underline{0.3232} & -- & \underline{0.5420} & --\\
& IntEL & \underline{0.8466} & 0.4685 & 0.3956 & -- & 0.6419 & -- & 0.5791 & 0.3074 & 0.3124 & -- & 0.5273 & --\\
& \textbf{UMRE} & \textbf{0.8763} & \textbf{0.5703} & \textbf{0.5012} & -- & \textbf{0.6854} & -- & \textbf{0.6012} & \textbf{0.3298} & \textbf{0.3431} & -- & \textbf{0.5485} & --\\
\hline
\end{tabular}
}
\label{tab:model_comparison}
\end{table*}

\begin{table*}[htbp]
\centering
\renewcommand{\arraystretch}{1.2} 
\caption{Experiment with the generalization of the UMNN module on the KuaiRand dataset and compare the NDCG@3 metrics at each task between the baseline model without the addition of UMNN and the model with the addition of UMNN.}
\label{tab:UMNN generalizability}
\resizebox{\textwidth}{!}{
\begin{tabular}{c|cccccc|cccccc}
\hline
\multirow{2}{*}{\diagbox{Model}{Metric}} & \multicolumn{6}{c|}{Base} & \multicolumn{6}{c}{+UMNN} \\
& Click & Like & Follow & Comment & Forward & Long view & Click & Like & Follow & Comment & Forward & Long view \\
\hline
SingleSort & 0.5529 & 0.1889 & 0.0534 & 0.0627 & 0.0580 & 0.4461 &0.5529 & 0.1889 & 0.0534 & 0.0627 & 0.0580 & 0.4461\\
LR         & 0.5841 & 0.1450 & 0.0437 & 0.0476 & 0.0363 & 0.5034 &0.6131 &0.1559 &0.0515 &0.0473 &0.0291 &0.5306\\
MLP        & 0.5731 & 0.1289 & 0.0612 & 0.0474 & 0.0384 & 0.4902 &0.6229 &0.2165 &0.1071 &0.1031 &0.0711 &0.5400\\
aWELv    & 0.5654 & 0.1063 & 0.0356 & 0.0652 & 0.0565 & 0.4622     & 0.5814     & 0.2020     & 0.1189     & 0.1025      & 0.0993     & 0.4852     \\
IntEL      & 0.6064 & 0.1255 & 0.0546 & 0.0422 & 0.0135 & 0.5197 &0.6628  &0.2310 &0.1353 &0.0421 &0.0549 &0.5795\\
\hline
\end{tabular}
}
\end{table*}

\subsubsection{Pre-training.}There is no pxtr prediction in public datasets thus we need to train a ranking model to obtain it, and since multi-task learning (MTL) is not the primary focus of this work, we employ the Progressive Layered Extraction (PLE) model to generate pxtr for each task. These predicted scores are then combined with the original dataset to serve as the input to our model, which outputs the final ensemble score through fusion of the pxtr.


\subsubsection{Baseline Methods.}
We compare UMRE with several representative ranking ensemble baselines:
\begin{itemize}
\item \textbf{Single Sort:} Directly ranks items by raw pxtr scores.
\item \textbf{LR:} A logistic regression model that linearly fuses pxtr with learnable weights.
\item \textbf{MLP:} A multi-layer perceptron that performs non-linear fusion of pxtr to generate final ranking scores.
\item \textbf{aWELv:} An ensemble learning framework that adaptively assigns weights to base models ~\cite{liu2022generalized}.
\item \textbf{IntEL:} An intent-aware ensemble model~\cite{li2023intent} that adaptively weights pxtr based on user intent.
\item \textbf{UMRE:} Our proposed an end-to-end personalized ranking ensemble model based on UMNN monotonic transformation.
\end{itemize}
For detailed descriptions, please refer to Appendix C.

\subsection{Performance of UMRE Models}
We compare several classical model-based ranking ensemble on two recommendation datasets, As shown in Table~\ref{tab:model_comparison}, we have the following findings:

(1) Our method UMRE outperforms compared to other ranking ensemble methods on two datasets, achieving the best performance on both evaluation metrics HR@3 and NDCG@3. The end-to-end ensemble training provides a larger space for parameter search, resulting in better performance.

(2) Compared with SingleSort, where the original pxtrs is evaluated on its corresponding task, our proposed UMRE also achieves better performance on all six tasks. MTL predicts the pxtr of multiple tasks at the same time, but it does not utilize the connection between the pxtr, and it only predicts the pxtr independently for a single task. There are associations between pxtrs, and some pxtrs are isotropic to each other. UMRE combines pxtr and user features as feature inputs, captures the connection between pxtr through self-attention, and obtains personalized ensemble scores after multi-task fusion to achieve better performance.

(3) The adaptive loss of Pareto optimality coordinates learning across multiple goals, compared to time-length tasks such as click, long view, where the behavior of interactive goals is sparser and thus more difficult to learn their predicted score. Pareto-optimal optimization strategy evaluates multiple objective training by boosting or suppressing the difference in metrics between two rounds of training to achieve Pareto-optimality for multi-tasks fusion. As shown in the experiments based on the kuairand dataset in Table~\ref{tab:model_comparison}, the UMRE with the addition of the Pareto-optimal optimization strategy has a very significant performance on the tasks where the original learning is poor.

\begin{table}[htbp]
\centering
\caption{Ablation study of UMRE on the KuaiRand dataset (NDCG@3 per task). Where $U_{\text{feat}}$ and $I_{\text{feat}}$ denote the User-side and Item-side features of the UMNN model inputs, respectively, and $Pareto_{\text{opt}}$ denotes the loss using Pareto Optimality.}
\label{tab:ablation_experiment}
{\small
\setlength{\tabcolsep}{2.5pt}  
\renewcommand{\arraystretch}{1.2}  
\begin{tabular}{c|ccccccc}
\hline
\textbf{Without} & Click & Like & Follow & Comment & Forward & Long view \\
\hline
- & \textbf{0.7712} & 0.3145 & \textbf{0.2693} & 0.1648 & \textbf{0.1393} & \textbf{0.7074} \\
$I_{\text{feat}}$              & 0.6576 & 0.1627 & 0.0633 & 0.0373 & 0.0237 & 0.5724 \\
$U_{\text{feat}}$              & 0.6919 & 0.2298 & 0.0925 & 0.1405 & 0.0515 & 0.6181 \\
$I_{\text{feat}},U_{\text{feat}}$ & 0.6308 & 0.1262 & 0.0226 & 0.0447 & 0.0513 & 0.5502 \\
$Pareto_{\text{opt}}$                       & 0.7695 & \textbf{0.3182} & 0.1870 & \textbf{0.1757} & 0.0650 & 0.7041 \\
\hline
\end{tabular}
}
\end{table}

\subsection{Performance of UMNN Transformation}

The UMNN module applies a pxtr-wise monotonic transformation that preserves the original ranking and can be integrated into any fusion model. To evaluate its generalization ability, we add UMNN to various baselines. As shown in Table~\ref{tab:UMNN generalizability}, all models show consistent gains in HR and NDCG, confirming the module’s effectiveness and compatibility.

UMNN is end-to-end trainable with multi-task fusion parameters, expanding the model’s capacity to fit complex functions while retaining the monotonicity constraint, which stabilizes convergence without disrupting pxtr order.

We further conduct ablation studies to analyze the impact of different input features. As shown in Table~\ref{tab:ablation_experiment}, even without feature input, UMNN improves performance by enlarging the parameter search space, albeit without personalized scaling. Adding user-side features enables user-specific transformations and yields further improvements. Incorporating item-side features brings the most significant gains, as it allows the model to apply item-sensitive adjustments while preserving monotonicity. Visualisation examples can be found in Appendix D.

\subsection{Performance of Pareto Optimality}

This section analyzes the impact of Pareto-optimal optimization strategy weights. We conduct ablation experiments to validate their effectiveness. Without the Pareto-optimal formulation, we must manually assign reward weights for training. To balance objectives, we set these weights in proportion to the positive sample rate of each task, assigning higher weights to tasks with less positive samples.

As shown in Table~\ref{tab:ablation_experiment}, the Pareto-optimal adaptive weighting consistently outperforms manually set rewards, especially for underperforming tasks. By dynamically identifying and emphasizing poorly optimized objectives during training, the Pareto mechanism improves their performance and ensures balanced multi-task optimization. 

\begin{table}[htbp]
\centering
\renewcommand{\arraystretch}{1.2}
\caption{The online performance of \textbf{UMRE}.}
\label{tab:online experiment}
\begin{tabular}{c|c}
\hline
 Task & Gains from the UMRE Online Experiment \\
\hline
Like      & \textbf{+5.477\%} \quad {[-0.41\%, 0.41\%]} \\
Follow          & \textbf{+2.730\%} \quad {[-0.62\%, 0.63\%]} \\
Forward           & \textbf{+5.023\%} \quad {[-0.66\%, 0.71\%]} \\
Comment          & \textbf{+6.408\%} \quad {[-0.58\%, 0.59\%]} \\
\hline
\end{tabular}
\end{table}
\subsection{Online Experiments}



We conducted an A/B test on a short video platform with over 400 million users to evaluate UMRE. The platform considers multiple interaction signals (like, follow, forward, comment), and we used the corresponding predicted scores (pltr, pwtr, pftr, pcmtr) as input pxtr for fusion.

To ensure fairness, users were randomly assigned into two groups, each receiving 5\% of real-time traffic. Prior to the experiment, interaction metrics across the groups were rebalanced to minimize distributional bias. The baseline adopts a formula-based MTF approach, and UMRE was tested during the fine ranking stage.

As shown in Table~\ref{tab:online experiment}, UMRE consistently outperforms the baseline across all interaction metrics, and the increase is significant on this platform, validating that personalized fusion leads to more interest-aligned recommendations in real-world scenarios.

\section{Conclusion}
In this work, we propose UMRE, an end-to-end monotonic transformed ranking ensemble model for multi-task fusion in recommender systems. UMRE leverages UMNN to apply a strictly monotonic transformation to the original pxtr scores, enabling effective fusion through the ranking model. By incorporating user features, it supports personalized fusion, while Pareto-optimal strategy ensures balanced multi-task optimization. We demonstrate the effectiveness of UMRE on two benchmark recommendation datasets and a large-scale online video platform.

\vspace{3em}
\bibliography{reference}

\appendix
\newpage
\clearpage
\section{A.  Detailed Implementation of UMNN Transformation}
\label{section:Detailed Implementation of UMNN Transformation}
\noindent \textbf{Mathematical Foundation}:  
Each transformation \(g_k(\cdot)\) is defined as:
\[
g_k(p_k, h; \theta_k) = \int_0^{p_k} f_k(t, h; \theta_k) \, dt + \beta_k
\]
where:
\begin{itemize}
    \item \(f_k(t, h; \theta_k)\) is an unconstrained neural network with strictly positive outputs, enforced via an ELU+1 activation;
    \item \(h\) is the feature vector containing the user embedding, item embedding;
    \item \(\theta_k\) denotes the parameters specific to the \(k\)-th transformation network;
    \item \(\beta_k\) is a learnable bias term.
\end{itemize}
The strict positivity constraint on \(f_k\) guarantees the monotonicity of \(g_k\), ensuring:
\[
p_k^{(a)} > p_k^{(b)} \implies g_k(p_k^{(a)}, h) > g_k(p_k^{(b)}, h)
\]
which is critical for preserving the relative order of prediction targets in ranking applications.

\noindent \textbf{Efficient Forward Computation}:  
During inference, the integral is approximated using Clenshaw-Curtis quadrature with \(Q\) nodes:
\[
t_k = g_k(p_k, h) \approx \sum_{q=1}^Q w_q f_k(t_q, h; \theta_k) + \beta_k
\]
where nodes \(\{t_q\}\) are precomputed based on the range of \(p_k\), and quadrature weights \(\{w_q\}\) ensure fast and accurate convergence. This approximation enables:
\begin{enumerate}
    \item Exponential convergence for smooth objective functions;
    \item Efficient batched evaluation across user-item pairs;
    \item Inference-time complexity independent of quadrature resolution.
\end{enumerate}

\noindent \textbf{Gradient-Based Optimization}:  
The gradient of the loss \(\mathcal{L}\) with respect to the network parameters \(\theta_k\) follows the Leibniz integral rule:
\[
\nabla_{\theta_k} \mathcal{L} = \frac{\partial \mathcal{L}}{\partial t_k} \left( \int_0^{p_k} \nabla_{\theta_k} f_k(t, h; \theta_k) \, dt \right)
\]
This formulation avoids storing intermediate integral results during backpropagation, yielding memory efficiency:
\[
\begin{aligned}
\mathcal{O}(\text{memory}) &\propto \text{model size}, \\
\text{vs.} 
\quad \mathcal{O}(Q) &\text{ in naive implementations}
\end{aligned}
\]

\noindent \textbf{Model Integration}:  
We instantiate a separate UMNN for each prediction target \(k\), each configured as follows:
\begin{enumerate}
    \item All networks \(f_k\) share the same architecture (3-layer MLP with residual connections), but have distinct parameters \(\theta_k\);
    \item Input to \(f_k\) includes pxtr score \(t\), combined context vector \(h\), and positional encoding;
    \item Output scores \(\mathbf{t} = (t_1, \dots, t_K)\) are passed to the attention-based fusion module.
\end{enumerate}

\begin{table}[h]
\centering
\caption{UMNN Specification}
\label{table:UMNN Specification}
\begin{tabular}{ll}
\hline
\textbf{Component} & \textbf{Configuration} \\
\hline
Integral approximation & Clenshaw-Curtis (Q=32) \\
Activation & ELU + 1 (output layer) \\
Network \(f_k\) & 3 $\times$ (Linear(128) + Relu) \\
Regularization & Weight decay ($\lambda=10^{-4}$) \\
Bias initialization & $\beta_k \sim \mathcal{U}(-0.5, 0.5)$ \\
\hline
\end{tabular}
\end{table}

This design equips our model with expressive, strictly monotonic transformations tailored to individual user and item characteristics, ensuring both interpretability and ranking reliability.

\section{B.  Detailed Description of Datasets}
\label{section:Detailed Description of Datasets}

The experiments are conducted on two publicly available datasets: the KuaiRand dataset for online short video recommendation, and the Tenrec dataset for online shopping recommendation. 

\textbf{KuaiRand} is a public dataset released by Kuaishou for short video recommendation, containing 27,285 users and 32,038,725 interaction records. It provides contextual features for both users and items, as well as a variety of user feedback signals. In this work, we consider six types of user interactions: click, long view, like, follow, comment, and share. To facilitate research and reduce computational complexity, we use the KuaiRand-1K subset, which randomly samples 1,000 users from the full dataset. The corresponding video items are also reduced to approximately 4 million, while preserving the richness and diversity of user behaviors.

\textbf{Tenrec} is Tencent's two infomercial (article and video) recommendation platforms, the QK platform and the QB platform, each of which has two feeds, i.e., articles and videos, with varying degrees of overlap ratios between the two platforms. We choose QK-video for our experiments, which contains four types of user feedback behaviors, namely click, follow, like, and forward (share), and due to its huge amount of data, we intercept the 50k users with the most number of exposures to perform the experiments. 

Original data sets consist of individual exposure records. To simulate candidate ranking in an online inference setting, we segment the data into sessions based on timestamps: interactions occurring within a 1-hour window are grouped into the same session. We filter out sessions with fewer than 20 interactions and, for each remaining session, we collect the user's historical interaction sequence prior to the session. The maximum length of this sequence is set to 30.

\section{C.  Detailed Description of Baseline Methods}
\subsubsection{Baseline Methods.}We compare UMRE against basic models and several ranking ensemble baselines as follows,
\begin{itemize}
    \item \textbf{Single Sort:} Sorted based on raw pxtr scores and evaluated using the pxtr sort results for the evaluation.
    \item \textbf{LR:} Logistic Regression model is a linear binary classification model via a sigmoid activation function that is used to predict the probability of positive and negative samples, here we use clicks as label. It weighted fusion of the input pxtr, which is essentially a summation formula with learnable weights.
    \item \textbf{MLP:} Multilayer perceptron is a feed forward artificial neural network that consists of multiple neurons (neural nodes) and has a deeper structure as compared to LR, which also uses pxtr as input for single-point fusion output fusion scores, and it also uses clicks for label training.
    \item \textbf{aWELv:} aWELv (adaptive Weighting Ensemble Learning via validation) is an ensemble learning framework that adaptively assigns weights to base models according to their validation performance. Instead of relying on fixed or manually tuned fusion coefficients, aWELv estimates the optimal combination by minimizing a validation loss function, often based on the final objective metric (e.g., AUC or NDCG).
    \item \textbf{IntEL:} This Intent-aware ranking Ensemble Learning model obtains $w_x(u,i)$ by learning the user intent and combining it with the input pxtr and the user category through the intent-aware ranking ensemble module. $w_x(u,i)$ denotes the weight of an item in the user's queue of a certain pxtr, and finally the ensemble score is obtained by additive fusion $S_i^{ens} =  {\textstyle \sum_{x}^{}}w_x(u,i)s_i^{x}$.
    \item \textbf{UMRE:} We propose an end-to-end personalized ranking ensemble model based on UMNN monotonic transformation.
\end{itemize}

\newpage
\section{D. UMNN Monotonic Transformation Visualisation}
\label{section:D}

To better illustrate the behavior of the UMNN module, we visualize its monotonic transformations on four representative user sessions. Each subplot compares the original \textit{pxtr} inputs with their UMNN-transformed outputs. The model learns personalized, nonlinear, yet strictly monotonic mappings that better align scoring with user-specific preferences.

\begin{figure}[htbp]
    \centering
    \setlength{\abovecaptionskip}{4pt}

    \includegraphics[width=\linewidth]{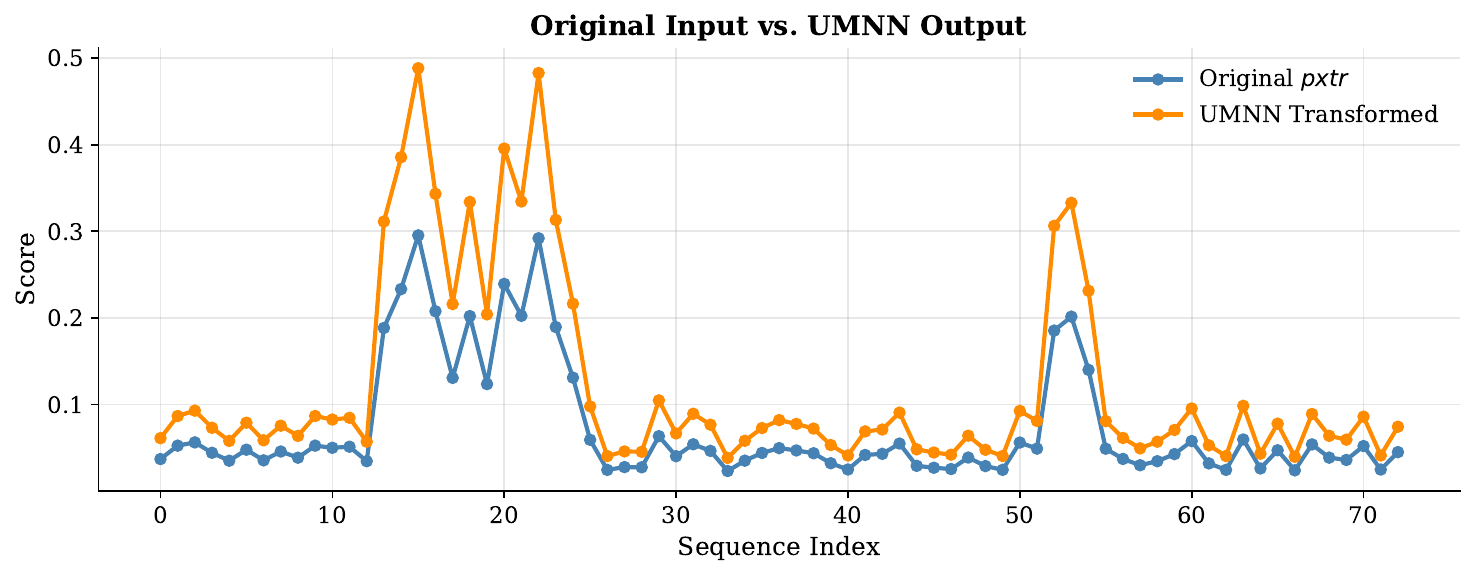}
    \caption*{(a) Session 1.}

    \includegraphics[width=\linewidth]{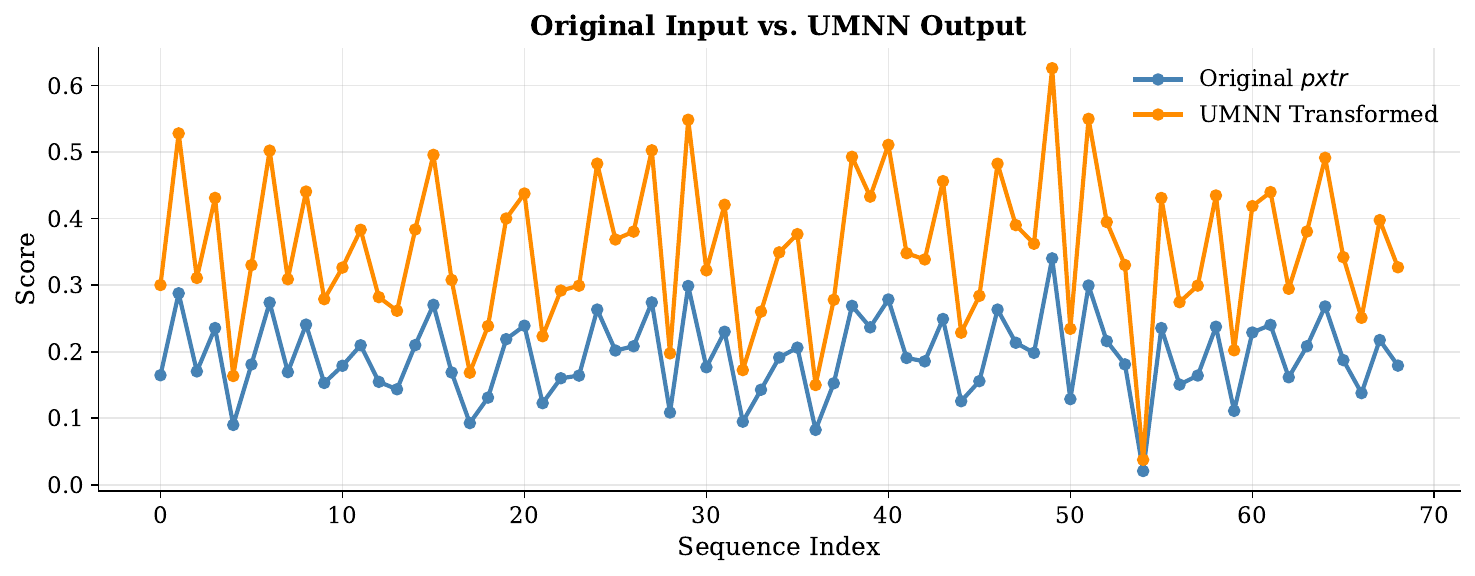}
    \caption*{(b) Session 2.}

    \includegraphics[width=\linewidth]{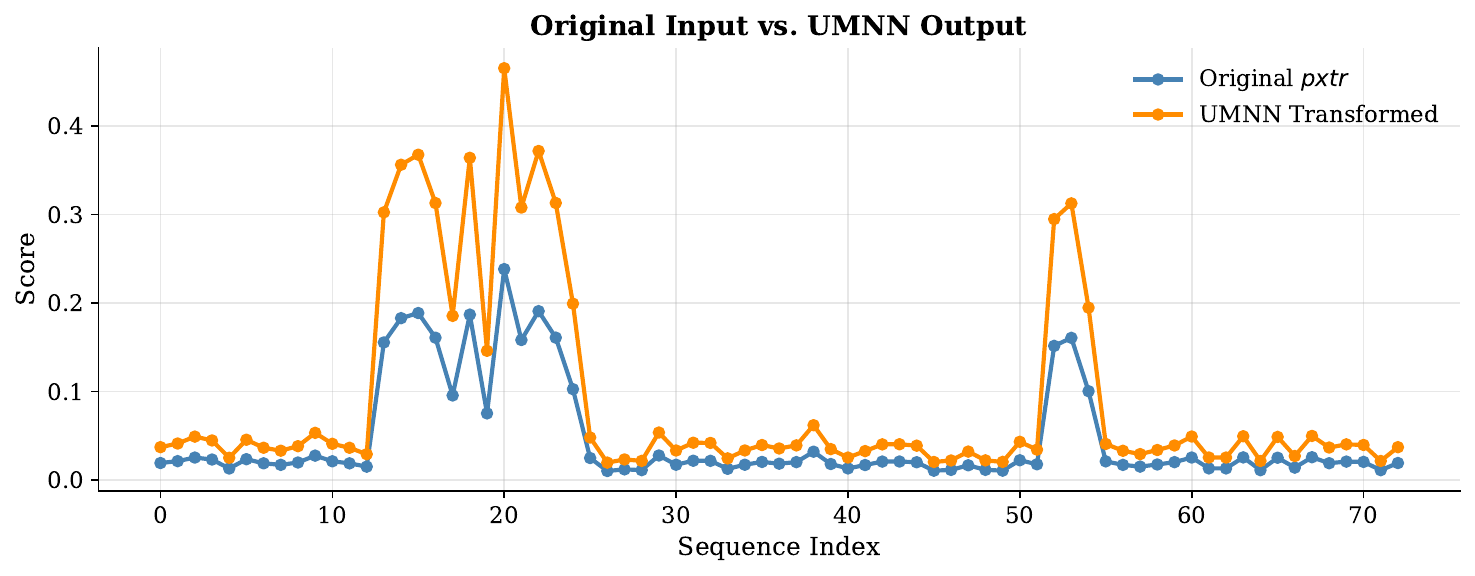}
    \caption*{(c) Session 3.}

    \includegraphics[width=\linewidth]{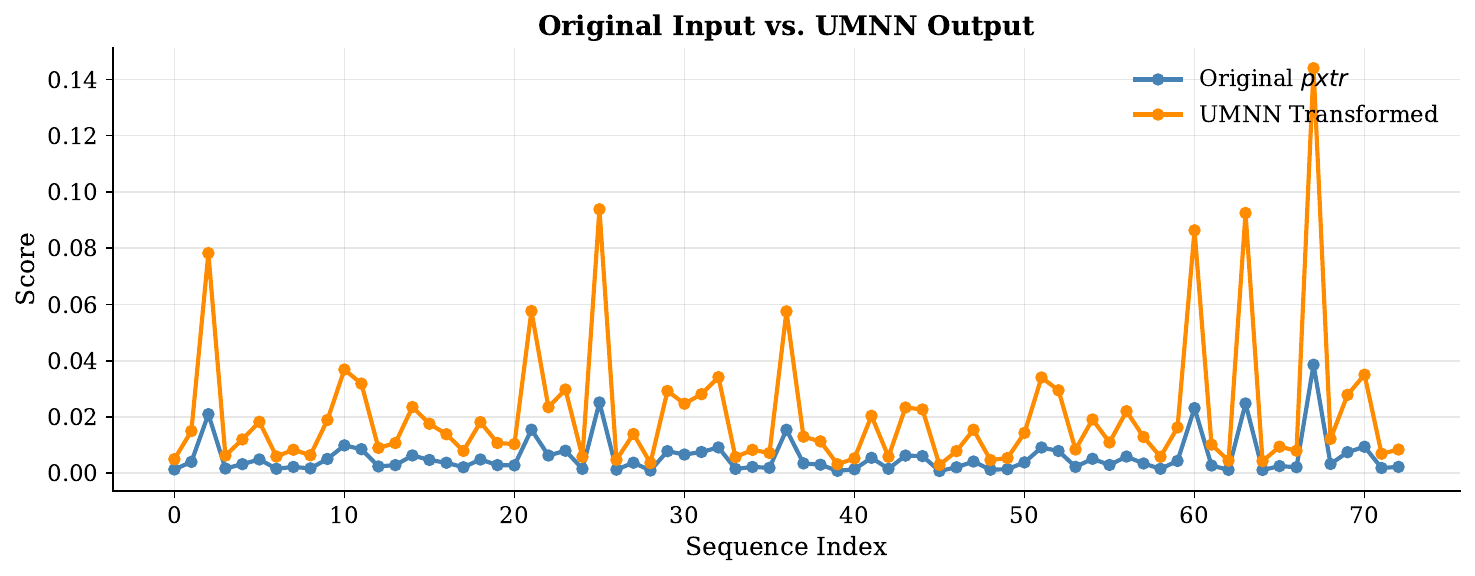}
    \caption*{(d) Session 4.}

    \caption{Visualization of UMNN-transformed \textit{pxtr} values across four user sessions. The learned transformations are smooth and strictly monotonic.}
    \label{fig:umnn_transform_all}
\end{figure}

\end{document}